# An Incoercible E-Voting Scheme Based on Revised Simplified Verifiable Re-encryption Mix-nets

Shinsuke Tamura[1,*], Hazim A. Haddad[1], Nazmul Islam[2], Kazi Md. Rokibul Alam[2]

[1]Graduate School of Engineering, University of Fukui, Japan
[2]Department of Computer Science and Engineering,
Khulna University of Engineering and Technology, Bangladesh
[*]Corresponding author: tamura@u-fukui.ac.jp

*Abstract*– Simplified verifiable re-encryption mix-net (SVRM) is revised and a scheme for e-voting systems is developed based on it. The developed scheme enables e-voting systems to satisfy all essential requirements of elections. Namely, they satisfy requirements about privacy, verifiability, fairness and robustness. It also successfully protects voters from coercers except cases where the coercers force voters to abstain from elections. In detail, voters can conceal correspondences between them and their votes, anyone can verify the accuracy of election results, and interim election results are concealed from any entity. About incoercibility, provided that erasable-state voting booths which disable voters to memorize complete information exchanged between them and election authorities for constructing votes are available, coercer *C* cannot know candidates that voters coerced by *C* had chosen even if the candidates are unique to the voters. In addition, elections can be completed without reelections even when votes were handled illegitimately.

*Keywords*– revised-SVRM, anonymous credential, inferior votes, erasable-state voting booth.

## 1. Introduction

E-voting systems are expected to make elections efficient, accurate and economical, but when elections are computerized, voters are faced with serious threats. For example, simple computerization enables election authorities to know correspondences between voters and their votes. Also, entity *C* that is coercing voter *V* becomes able to confirm whether *V* had chosen *C*'s designating candidate or not. E-voting systems applicable to real elections must satisfy the following requirements.

1. *Privacy* Correspondences between voters and their votes must be concealed from others including election authorities. It is preferable that voters can conceal also their abstentions from others.

2. *Verifiability* Anyone including voters and third parties must be able to verify the accuracy of elections, i.e. e-voting schemes must be able to convince anyone that only and all votes from eligible voters had been counted.

3. *Fairness* Interim election results influence ways voters choose candidates; therefore interim election results must be concealed from anyone including election authorities.

4. *Incoercibility* To disable entity *C* that is coercing voter *V* to confirm that *V* actually had chosen *C*'s designating candidate *S*, e-voting schemes must disable even *V* itself to identify its vote in election results. Here, *C* must be disabled to know whether *V* had chosen *S* or not even if *S* is unique to *V*.

5. *Robustness* To conduct elections fairly even when relevant entities behave dishonestly, voting schemes must be able to complete elections without reelections or any help of dishonest voters. Here if helps from dishonest voters are required, the schemes cannot complete the election when they disappear.

However, despite that many schemes had been developed, they cannot satisfy all of the above requirements completely [2,4,6,8,10,11,12]. For example, although many schemes satisfy receipt freeness, if entity *C* that is coercing voter *V* asks *V* to choose candidate *S* that is unique to *V*, *C* can easily know whether *V* had actually chosen *S* or not. Here, receipt freeness is the base of incoercibility, i.e. it disables *C* to force *V* to show its receipt that includes the candidate *V* had chosen.

This paper modifies simplified verifiable re-encryption mix-net (SVRM) [14] to revised-SVRM, and develops an e-voting scheme that satisfies all the above requirements based on it together with anonymous tag based credentials [13,15,16]. An anonymous credential enables voter *V* to convince others that it is eligible without revealing its identity, and provided that erasable-state voting booths are available, the verifiable feature of revised-SVRM ensures election authorities' legitimate handling of votes while concealing correspondences between voters and their votes from entities including voters themselves. In addition, the scheme regards votes for candidates that could not obtain enough supports as inferior votes and does not count them in the tallying phase [13]. Then, entity *C* that is coercing *V* cannot confirm whether *V* had chosen *C*'s designating candidate or not even when the–candidate is unique to V. About fairness and robustness, it is easy to satisfy them as same as in other schemes.

In the above, an erasable-state voting booth is a one that disables voters to memorize complete information exchanged between them and election authorities during they are constructing their votes.

## 2. Security Components

### 2.1. Simplified Verifiable Re-encryption Mix-net (SVRM)

Re-encryption mix-net $M$ consists of multiple mutually independent mix-servers $M_1, M_2, ---, M_Q$ and $M_Q, M_{Q-1}, ---, M_1$ that are arrayed in encryption and decryption stages respectively as shown in Figure 1. Then, $M$ enables entities $V_1, V_2, ---, V_N$ that put their attribute values $D_1, D_2, ---, D_N$ in it to conceal correspondences between them and $D_1, D_2, ---, D_N$ from others including mix-servers [2-7].

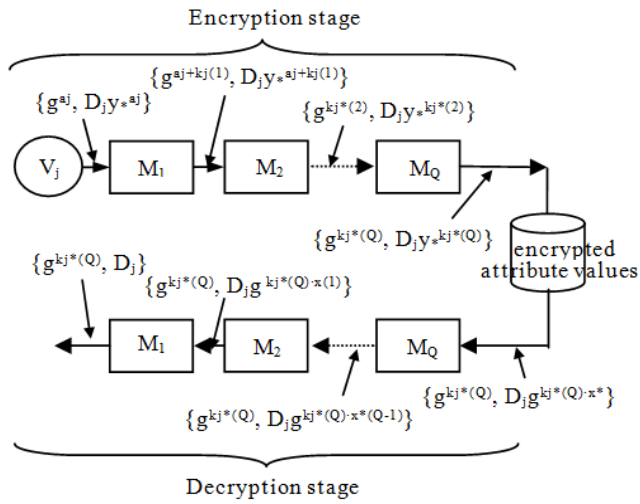

**Figure 1.** Re-encryption mix-net $M$

To conceal the above correspondences, firstly each $V_j$ encrypts its attribute value $D_j$ to $\{g^{a_j}_{\bmod p}, D_jy_*^{a_j}{}_{\bmod p}\}$ while using its secret integer $a_j$, and $M_1, M_2, ---, M_Q$ in the encryption stage repeatedly encrypt $\{g^{a_j}_{\bmod p}, D_jy_*^{a_j}{}_{\bmod p}\}$ to $\{g^{k_j^*(Q)}{}_{\bmod p}, D_jy_*^{k_j^*(Q)}{}_{\bmod p}\}$ by using their secret integers $k_j(1), k_j(2), ---, k_j(Q)$ to be decrypted in the decryption stage. Here, provided that g and p are publicly known appropriate integers (p is a prime number), $\{x(q), y(q) = g^{x(q)}{}_{\bmod p}\}$ is mix-server $M_q$'s secret decryption and public encryption key pair of an ElGamal encryption function, and $<x_* = \{x(1)+x(2)+ --- +x(Q)\}_{\bmod p}, y_* = y(1)y(2)---y(Q)_{\bmod p} = g^{x^*}{}_{\bmod p}>$ is a common secret decryption and public encryption key pair. Also, $k_j^*(q) = a_j+k_j(1)+k_j(2)+ --- +k_j(q)$, and in the remainder, notation $_{\bmod p}$ is omitted when confusions can be avoided.

In detail, each $M_q$ in the encryption stage calculates $\{g^{k_j^*(q-1)}g^{k_j(q)} = g^{k_j^*(q)}, D_jy_*^{k_j^*(q-1)}y_*^{k_j(q)} = D_jy_*^{k_j^*(q)}\}$ from $\{g^{k_j^*(q-1)}, D_jy_*^{k_j^*(q-1)}\}$ received from $M_{q-1}$. After that $M_q$ shuffles its calculation results, and forwards them to $M_{q+1}$. In the decryption stage, $M_Q, M_{Q-1}, ---, M_1$ decrypt each $\{g^{k_j^*(Q)}, D_jy_*^{k_j^*(Q)} = D_jg^{k_j^*(Q)\cdot x^*}\}$ to $\{g^{k_j^*(Q)}, D_j\}$. Namely, each $M_q$ decrypts $\{g^{k_j^*(Q)}, D_jg^{k_j^*(Q)\cdot x^*(q)}\}$ received from $M_{q+1}$ to $\{g^{k_j^*(Q)}, D_jg^{k_j^*(Q)\cdot x^*(q)}/g^{k_j^*(Q)\cdot x(q)} = D_jg^{k_j^*(Q)\cdot x^*(q-1)}\}$ by using its secret key $x(q)$, and forwards it to $M_{q-1}$. Here $x^*(q) = x(1)+x(2)+ --- +x(q)$.

Then, no one except $V_j$ can know $V_j$'s attribute value $D_j$ unless all mix-servers conspire because any one cannot know all integers $k_j(1), k_j(2), ---, k_j(Q)$ or all decryption keys $x(1), x(2), ---, x(Q)$. Entities other than $V_j$ cannot know integer $a_j$ either.

However, because integers $k_1(q), k_2(q), ---, k_N(q)$ and decryption key $x(q)$ are known only to $M_q$ and $M_q$ in the encryption stage shuffles its encryption results, no one can notice even when mix-servers encrypt or decrypt attribute values dishonestly. SVRM $M_*$ shown in Figure 2 enables any entity $E$ to verify behaviors of mix-servers by preparing the unknown number generation stage [14]. In the following it is assumed that all information sent from each entity $V_j$ and mix-server $M_q$ is publicly disclosed.

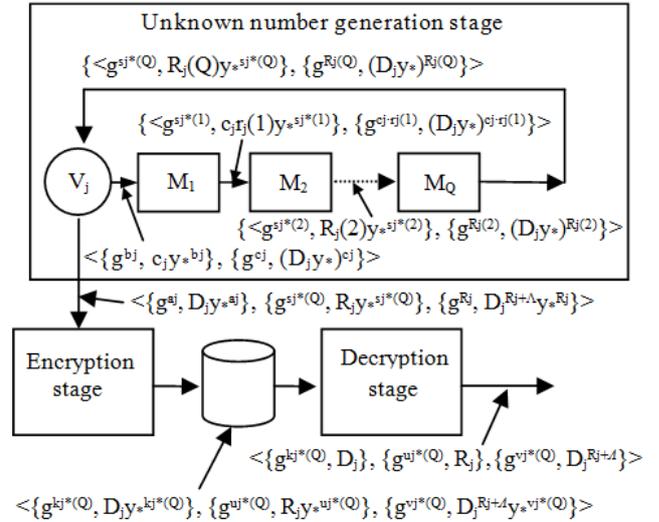

**Figure 2.** Simplified verifiable re-encryption mix-net $M_*$

Firstly, provided that $b_j, c_j$ are integers secrets of entity $V_j$ and $r_j(q)$ and $s_j(q)$ are integers secrets of each mix server $M_q$, $V_j$ calculates $\{g^{b_j}, c_jy_*^{b_j}\}$ and $\{g^{c_j}, (D_jy_*)^{c_j}\}$, and forwards them to $M_1$ in the unknown number generation stage. After that each $M_q$ calculates $<\{g^{s_j^*(q-1)}g^{s_j(q)} = g^{s_j^*(q)}, R_j(q-1)r_j(q)y_*^{s_j^*(q-1)}y_*^{s_j(q)} = R_j(q)y_*^{s_j^*(q)}\}, \{g^{R_j(q-1)\cdot r_j(q)} = g^{R_j(q)}, (D_jy_*)^{R_j(q-1)\cdot r_j(q)} = (D_jy_*)^{R_j(q)}\}>$ from $<\{g^{s_j^*(q-1)}, R_j(q-1)y_*^{s_j^*(q-1)}\}, \{g^{R_j(q-1)}, (D_jy_*)^{R_j(q-1)}\}>$ calculated by $M_{q-1}$. As a result, finally $M_Q$ calculates $<\{g^{s_j^*(Q)}, R_jy_*^{s_j^*(Q)}\}, \{g^{R_j}, (D_jy_*)^{R_j}\}>$ to forward it to $V_j$, and $V_j$ constructs triplet $E_0(D_j) = <\{g^{a_j}, D_jy_*^{a_j}\}, \{g^{s_j^*(Q)}, R_jy_*^{s_j^*(Q)}\}, \{g^{R_j}, D_j{}^{\Lambda}D_j{}^{R_j}y_*^{R_j} = D_j{}^{R_j+\Lambda}y_*^{R_j}\}>$ to put in mix-net $M_*$. Where, $\Lambda$ is a publicly known constant integer, $s_j^*(q) = b_j+s_j(1)+s_j(2)+ --- +s_j(q), R_j(q) = c_jr_j(1)r_j(2) --- r_j(q)$ and $R_j = R_j(Q)$.

About mix-servers $M_1, M_2, ---, M_Q$ in the encryption stage, they repeatedly encrypt $E_0(D_j)$ to triplet $E_Q(D_j) = <\{g^{k_j^*(Q)}, D_jy_*^{k_j^*(Q)}\}, \{g^{u_j^*(Q)}, R_jy_*^{u_j^*(Q)}\}, \{g^{v_j^*(Q)}, D_j{}^{R_j+\Lambda}y_*^{v_j^*(Q)}\}>$ while shuffling its all encryption results as same as in Figure 1, and $M_Q, M_{Q-1}, ---, M_1$ in the decryption stage decrypt it to $F_0(D_j) = <\{g^{k_j^*(Q)}, D_j\}, \{g^{u_j^*(Q)}, R_j\}, \{g^{v_j^*(Q)}, D_j{}^{R_j+\Lambda}\}>$. Namely, $M_q$ in the encryption stage calculates $E_q(D_j) = <\{g^{k_j^*(q)}, D_jy_*^{k_j^*(q)}\}, \{g^{u_j^*(q)}, R_jy_*^{u_j^*(q)}\}, \{g^{v_j^*(q)}, D_j{}^{R_j+\Lambda}y_*^{v_j^*(q)}\}>$ from $E_{q-1}(D_j) = <\{g^{k_j^*(q-1)}, D_jy_*^{k_j^*(q-1)}\}, \{g^{u_j^*(q-1)}, R_jy_*^{u_j^*(q-1)}\}, \{g^{v_j^*(q-1)}, D_j{}^{R_j+\Lambda}y_*^{v_j^*(q-1)}\}>$ calculated by $M_{q-1}$, and $M_q$ in the decryption stage calculates $F_{q-1}(D_j) = <\{g^{k_j^*(Q)}, D_jg^{k_j^*(Q)\cdot x^*(q-1)}\}, \{g^{u_j^*(Q)}, R_jg^{u_j^*(Q)\cdot x^*(q-1)}\}, \{g^{v_j^*(Q)}, D_j{}^{R_j+\Lambda}g^{v_j^*(Q)\cdot x^*(q-1)}\}>$ from $F_q(D_j) = <\{g^{k_j^*(Q)}, D_jg^{k_j^*(Q)\cdot x^*(q)}\}, \{g^{u_j^*(Q)}, R_jg^{u_j^*(Q)\cdot x^*(q)}\}, \{g^{v_j^*(Q)}, D_j{}^{R_j+\Lambda}g^{v_j^*(Q)\cdot x^*(q)}\}>$ calculated by $M_{q+1}$. Where, $u_j(q)$ and $v_j(q)$ are secret integers of $M_q$ and $u_j^*(q) = s_j^*(Q) + u_j(1) + u_j(2) + \cdots + u_j(q)$ and $v_j^*(q) = R_j + v_j(1) + v_j(2) + \cdots + v_j(q)$.

Then, the final decryption results enable $M_*$ to convince any verifier $E$ of its legitimate encryptions, decryptions and shuffling. Namely, because each attribute value $D_j$ is finally decrypted to $F_0(D_j) = <\{g^{kj*(Q)}, D_j = \alpha_j\}, \{g^{uj*(Q)}, R_j = \beta_j\}, \{g^{vj*(Q)}, D_j^{Rj+\Lambda} = \gamma_j\}>$, at least one mix-server is dishonest when relation $\alpha_j^{\beta j+\Lambda} = \gamma_j$ does not hold for some j.

However, each $M_q$ that knows public encryption keys $y(1), y(2), \cdots, y(Q)$ can easily forge encryption and decryption forms $E_q(D_j)$ and $F_{q-1}(D_j)$ so that their final decryption result $<\{g^{kj*(Q)}, \alpha_j\}, \{g^{uj*(Q)}, \beta_j\}, \{g^{vj*(Q)}, \gamma_j\}>$ satisfies relation $\alpha_j^{\beta j+\Lambda} = \gamma_j$. $M_*$ removes this possibility as below.

Firstly each $M_q$ discloses $\kappa_q = k_1(q) + k_2(q) + \cdots + k_N(q)$, and verifier $E$ convinces itself that the product of $E_q(D_1)$, $E_q(D_2), \cdots, E_q(D_N)$ calculated by $M_q$ and that of $E_{q-1}(D_1)$, $E_{q-1}(D_2), \cdots, E_{q-1}(D_N)$ calculated by $M_{q-1}$ are consistent. In detail, $E$ examines relations $G_{k1}(q) = g^{\kappa q}G_{k1}(q-1)$ and $G_{k2}(q) = y_*^{\kappa q}G_{k2}(q-1)$, and requests $M_q$ to iterate the encryption stage until the relations hold. Where $\{G_{k1}(q), G_{k2}(q)\}$ is a product pair $\{G_{k1}(q) = g^{k1*(q)}g^{k2*(q)} \cdots g^{kN*(q)} = g^{\kappa 1+\kappa 2+ \cdots +\kappa q}, G_{k2}(q) = D_1y_*^{k1*(q)}D_2y_*^{k2*(q)} \cdots D_Ny_*^{kN*(q)} = D_1D_2 \cdots D_Ny_*^{\kappa 1+\kappa 2+ \cdots +\kappa q}\}$. Therefore $G_{k1}(q) = g^{\kappa q}G_{k1}(q-1)$ and $G_{k2}(q) = y_*^{\kappa q}G_{k2}(q-1)$ necessarily hold if $E_q(D_1)$, $E_q(D_2), \cdots, E_q(D_N)$ are correct. But if encryption from $E_q(D_j)$ is incorrect, because solving discrete logarithm problems is difficult, $M_q$ that does not know $x_*$ cannot find value $\kappa_q$ so that $G_{k1}(q) = g^{\kappa q}G_{k1}(q-1)$ and $G_{k2}(q) = y_*^{\kappa q}G_{k2}(q-1)$ hold [3,4]. On the other hand, although $M_q$ discloses $\kappa_q$, it can maintain each $k_j(q)$ as its secret.

Here, actually $M_q$ can find integer $\kappa_q$ even if $E_q(D_j)$ is incorrect when $E_q(D_j)$ is calculated in a specific way, but in this case final decryption result $<\{g^{kj*(Q)}, \alpha_j\}, \{g^{uj*(Q)}, \beta_j\}, \{g^{vj*(Q)}, \gamma_j\}>$ does not satisfy relation $\alpha_j^{\beta j+\Lambda} = \gamma_j$. For example, if $M_q$ encrypts $\{g^{kj*(q-1)}, D_jy_*^{kj*(q-1)}\}$ in $E_{q-1}(D_j)$ and $\{g^{kh*(q-1)}, D_hy_*^{kh*(q-1)}\}$ in $E_{q-1}(D_h)$ to $\{g^{kj*(q)}, \lambda D_jy_*^{kj*(q)}\}$ and $\{g^{kh*(q)}, (1/\lambda)D_hy_*^{kh*(q)}\}$ instead of $\{g^{kj*(q)}, D_jy_*^{kj*(q)}\}$ and $\{g^{kh*(q)}, D_hy_*^{kh*(q)}\}$ ($\lambda$ is an arbitrarily integer), value $\kappa_q = k_1(q) + k_2(q) + \cdots + k_N(q)$ still satisfies $G_{k1}(q) = g^{\kappa q}G_{k1}(q-1)$ and $G_{k2}(q) = y_*^{\kappa q}G_{k2}(q-1)$. However, $M_q$ that does not know $D_j$, $R_j$, $D_h$ or $R_h$ cannot calculate $\{g^{uj*(q)}, \beta_jy_*^{uj*(q)}\}$, $\{g^{vj*(q)}, \gamma_jy_*^{vj*(q)}\}$ in $E_q(D_j)$ or $\{g^{uh*(q)}, \beta_hy_*^{uh*(q)}\}$, $\{g^{vh*(q)}, \gamma_hy_*^{vh*(q)}\}$ in $E_q(D_h)$ so that relations $\gamma_j = (\lambda D_j)^{\beta j+\Lambda}$ and $\gamma_h = (D_h/\lambda)^{\beta j+\Lambda}$ hold.

Secondly to ensure legitimate behaviors of mix-servers in the decryption stage, after $M_1$ having decrypted all attribute values, verifier $E$ calculates products $D_1D_2 \cdots D_N$ and $y_*^{a1+a2+ \cdots +aN}$ from $\{g^{k1*(Q)}, D_1\}, \cdots, \{g^{kN*(Q)}, D_N\}$ in final decryption results $F_0(D_1), \cdots, F_0(D_N)$ and $\{g^{a1}, D_1y_*^{a1}\}, \cdots, \{g^{aN}, D_Ny_*^{aN}\}$ in initial encryption forms $E_0(D_1), \cdots, E_0(D_N)$, where $E$ can calculate $y_*^{a1+a2+ \cdots +aN}$ as $y_*^{a1+a2+ \cdots +aN} = D_1y_*^{a1}D_2y_*^{a2} \cdots D_Ny_*^{aN}/(D_1D_2 \cdots D_N)$. $E$ calculates also $G_d$ from encryption forms $E_Q(D_1), E_Q(D_2), \cdots, E_Q(D_N)$ as $G_d = (D_1y_*^{k1*(Q)}D_2y_*^{k2*(Q)} \cdots D_Ny_*^{kN*(Q)})/(D_1D_2 \cdots D_N)$.

Under these settings, $E$ determines mix-servers in the decryption stage are dishonest when relation $G_d = y_*^{a1+a2+ \cdots +aN+\kappa 1+\kappa 2+ \cdots +\kappa Q}$ does not hold. Namely, apparently $G_d$ must be equal to $y_*^{a1+a2+ \cdots +aN+\kappa 1+\kappa 2+ \cdots +\kappa Q}$ if $E_Q(D_1), \cdots, E_Q(D_N)$ are correctly decrypted. On the other hand, it is computationally infeasible to find different values that satisfy relation $G_d = y_*^{a1+a2+ \cdots +aN+\kappa 1+\kappa 2+ \cdots +\kappa Q}$ as the decryption forms. In detail, although $M_q$ can forge $F_{q-1}(D_j)$ while satisfying $G_d = y_*^{a1+a2+ \cdots +aN+\kappa 1+\kappa 2+ \cdots +\kappa Q}$ as $M_q$ in the encryption stage calculates $\{g^{kj*(q)}, \lambda D_jy_*^{kj*(q)}\}$, it cannot make final decryption form $<\{g^{kj*(Q)}, \alpha_j\}, \{g^{uj*(Q)}, \beta_j\}, \{g^{vj*(Q)}, \gamma_j\}>$ satisfy relation $\alpha_j^{\beta j+\Lambda} = \gamma_j$ because each $M_q$ ($q > 1$) does not know value $D_j$ at a time when it decrypts $F_q(D_j)$. Also, anyone can examine relation $G_d = y_*^{a1+a2+ \cdots +aN+\kappa 1+\kappa 2+ \cdots +\kappa Q}$ because $y_*$ and $\kappa_1, \kappa_2, \cdots, \kappa_Q$ are publicly known.

In the above, mix-server $M_1$ in the decryption stage which calculates final decryption forms can know $D_1, \cdots, D_N$ as an exception. But verifier $E$ can detect dishonesties easily if the liable mix-server is $M_1$. Namely, when mix-servers $M_L, M_{L-1}, \cdots, M_1$ ($L < Q$) disclose their decryption keys $x(L), x(L-1), \cdots, x(1)$ after all attribute values were decrypted, verification of $M_1$'s behavior is trivial. Nevertheless, correspondences between entities and their attribute values can be concealed because $x(L+1), x(L+2), \cdots, x(Q)$ are still secrets of $M_{L+1}, M_{L+2}, \cdots, M_Q$.

Then, encryption and decryption forms $E_q(D_1), E_q(D_2), \cdots, E_q(D_N)$ and $F_{q-1}(D_1), F_{q-1}(D_2), \cdots, F_{q-1}(D_N)$ calculated by $M_q$ necessarily satisfy $G_{k1}(q) = g^{\kappa q}G_{k1}(q-1)$ and $G_{k2}(q) = y_*^{\kappa q}G_{k2}(q-1)$ for each q and $G_d = y_*^{a1+a2+ \cdots +aN+\kappa 1+\kappa 2+ \cdots +\kappa Q}$, and $E$ becomes able to detect illegitimately calculated $F_0(D_j) = <\{g^{kj*(Q)}, \alpha_j\}, \{g^{tj*(Q)}, \beta_j\}, \{g^{vj*(Q)}, \gamma_j\}>$ as the violation of relation $\alpha_j^{\beta j+\Lambda} = \gamma_j$.

About entities that are liable for inconsistent decryption results, verifier $E$ identifies them by tracing inconsistent decryption results back to initial encryption forms individually. In detail, provided that $F_0(D_j)$ is inconsistent, firstly $E$ asks $M_1$ in the decryption stage to show $F_1(D_j)$ from which it had calculated $F_0(D_j)$ and to prove correct calculation of $F_0(D_j)$. In the same way, $E$ asks each $M_q$ to show $F_q(D_j)$ from which it calculated $F_{q-1}(D_j)$ and to prove correct calculation of $F_{q-1}(D_j)$, and determines $M_q$ that cannot show consistent pair $\{F_q(D_j), F_{q-1}(D_j)\}$ is dishonest. $E$ asks each $M_q$ also in the encryption stage to show $E_{q-1}(D_j)$ from which it had calculated $E_q(D_j)$ and to prove correct calculation of $E_q(D_j)$. Then it determines $M_q$ is dishonest when $M_q$ cannot show consistent pair $\{E_{q-1}(D_j), E_q(D_j)\}$.

Here, $E$ can verify the consistency of $\{F_q(D_j), F_{q-1}(D_j)\}$, i.e. consistency between $\{g^{kj*(Q)}, D_jg^{kj*(Q) \cdot x*(q)}\}$ and $\{g^{kj*(Q)}, D_jg^{kj*(Q) \cdot x*(q-1)}\}$ without knowing secret key $x(q)$ by exploiting the scheme of Diffie and Hellman. Firstly, $E$ generates secret integer $\Psi$, and calculates $g^{kj*(Q) \cdot \Psi} = G_*$ and $\{D_jg^{kj*(Q) \cdot x*(q)}/D_jg^{kj*(Q) \cdot x*(q-1)}\}^{\Psi} = \{g^{kj*(Q) \cdot x(q)}\}^{\Psi}$. After that it asks $M_q$ to calculate $G_*^{x(q)}$ by showing $G_*$, and determines $M_q$ is dishonest when $G_*^{x(q)}$ is not equal to $\{D_jg^{kj*(Q) \cdot x*(q)}/D_jg^{kj*(Q) \cdot x*(q-1)}\}^{\Psi}$.

Verification of $\{E_{q-1}(D_j), E_q(D_j)\}$ is trivial, i.e. $M_q$ can disclose integer $k_j(q)$, because its value is changed at every encryption different from secret key $x(q)$. Also, although $E$ must verify behaviors of mix-servers in the unknown number generation stage if mix-servers in the encryption and decryption stages are honest, these verifications are trivial. As same as in the encryption stage each $M_q$ can disclose its secret integers $s_j(q)$ and $r_j(q)$.

Provided that a dishonest mix-server is not $M_1$ in the encryption stage or a mix-server in the unknown number generation stage, it is also straightforward to recalculate consistent final decryption result $F_0(D_j)$ without knowing corresponding entity $V_j$. But, when $M_1$ in the encryption stage or a mix-server in the unknown number generation stage is dishonest, entities other than $V_j$ may know the correspondence between $V_j$ and $D_j$, i.e. $E_Q(D_j)$ was decrypted already and the above procedure for identifying dishonest mix-servers reaches $E_0(D_j)$ that was put by $V_j$. By the same reason, $V_j$ cannot maintain $D_j$ as its secret when it put $D_j$ illegitimately. $M_*$ removes these threats by making $V_j$ anonymous. In addition about the latter threat, $V_j$ itself is responsible for the disclosure of its secrets.

Finally, it must be noted that because $y_*$, $g^{a1+\cdots+aN}$ and $\kappa_1, \kappa_2, \cdots, \kappa_Q$ are publicly known, any entity can confirm correct behaviors of SVRM without communicating with mix-servers. Therefore, although Diffie and Hellman scheme that requires interactions between a verifier and mix-servers is necessary to identify dishonest mix-servers, actual efficiency of SVRM is not degraded. Usually mix-servers are honest, i.e. they cannot continue their businesses once their dishonesties are detected.

## 2.2. Anonymous Tag Based Credential

Provided that $A$ is an authority that issues credentials and $Z$ is a secret integer of entity $V$, anonymous tag based credential $T(A, V, Z)$ enables $V$ to show its eligibility to any entity $E$ without revealing its identity. In addition, $E$ can force $V$ to calculate used seal $U^Z_{\mathrm{mod}\, B}$ from given integer $U$ by using integer $Z$ in $T(A, V, Z)$ honestly without knowing $Z$ itself ($B$ is a publicly known appropriate integer associated with $T(A, V, Z)$, and notation $_{\mathrm{mod}\, B}$ is omitted in the following). Then, $E$ can use $U^Z$ as an evidence that $V$ had shown $T(A, V, Z)$ to it. Here, actually $V$ shows $T(A, V, Z)$ to $E$ in form $T(A, V, Z)^W$ while generating secret integer $W$. Also, to maintain uniqueness of used seal $U^Z$, $V$ calculates a set of values $U_1^Z, U_2^Z, \cdots, U_T^Z$ from multiple integers $U_1, U_2, \cdots, U_T$.

In conclusion, together with used seal $U^Z$ anonymous credential $T(A, V, Z)$ satisfies unforgeability, soundness, anonymity, unlinkability, revocability and verifiability as below [13,15,16].

*Unforgeability* no one other than $A$ can generate valid credentials,

*Soundness* entities that do not know $Z$ in $T(A, V, Z)$ cannot prove the ownership of $T(A, V, Z)$ to other entity $E$. In addition, when $E$ illegitimately accepts $T(A, V, Z)$ shown by other entity $V_*$ possibly while conspiring with it, $A$ can detect that and identify liable entities,

*Anonymity* anyone except $V$ cannot identify $V$ from $T(A, V, Z)^W$ shown by $V$,

*Unlinkability* even if $V$ shows $T(A, V, Z)$ n-times in forms $T(A, V, Z)^{W1}, T(A, V, Z)^{W2}, \cdots, T(A, V, Z)^{Wn}$ while generating different secret integers $W_1, W_2, \cdots, W_n$, no one except $V$ can know links between them,

*Revocability* $A$ can invalidate $T(A, V, Z)$ without knowing secrets of honest entities, when its holder $V$ behaved dishonestly while showing $T(A, V, Z)^W$ or when $A$ reissued new credential to $V$ as a replacement of $T(A, V, Z)$, and

*Verifiability* anyone can verify the validity of $T(A, V, Z)$, in other words, entities can verify the validity without knowing any secret of $A$.

## 3. Revised-SVRM

To adapt SVRM to e-voting systems this section modifies it to revised-SVRM. Provided that $V_j$ and $D_j$ in Figure 2 are a voter and its vote respectively, SVRM cannot protect $V_j$ from coercer $C$, who is forcing $V_j$ to choose $C$'s designating candidate $S$. Namely, $V_j$ must disclose integers $b_j$, $c_j$ and pair $<\{g^{bj}, c_jy_*^{bj}\}, \{g^{cj}, (D_jy_*)^{cj}\}>$ that it had put in the unknown number generation stage in Figure 2 when $C$ requests. Then, $C$ can know whether $V_j$ actually had chosen $S$ or not by examining the consistency between $<\{g^{bj}, c_jy_*^{bj}\}, \{g^{cj}, (D_jy_*)^{cj}\}>$ and $S$. In the same way, $C$ can confirm $V_j$'s choice also from $\{g^{aj}, D_jy_*^{aj}\}$ in $E_0(D_j)$.

### 3.1. Modified Unknown Number Generation Stage

To disable entities to force $V_j$ to reveal attribute value $D_j$, revised-SVRM modifies the unknown number generation stage as shown in Figure 3. Here, as same as in Figure 2, although there is an exception information sent from mix-servers and each entity $V_j$ is publicly disclosed also in revised-SVRM. The modified unknown number generation stage proceeds as follow.

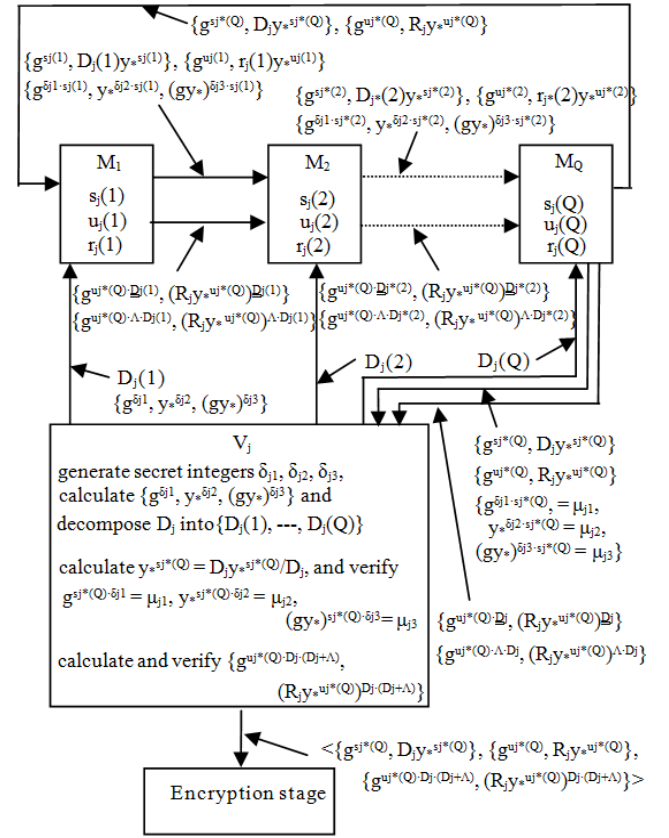

**Figure 3.** Modified unknown random number generation stage

1. Each mix-server $M_q$ in the unknown number generation stage generates its secret integers $s_j(q)$, $u_j(q)$ and $r_j(q)$.
2. Each entity $V_j$ generates its secret integers $\delta_{j1}, \delta_{j2}, \delta_{j3}$, calculates $\{g^{\delta j1}, y_*^{\delta j2}, (gy_*)^{\delta j3}\}$, and sends them to $M_1$. At the same time $V_j$ decomposes its attribute value $D_j$ into a set of values $\{D_j(1), D_j(2), \cdots, D_j(Q)\}$ so that product $D_j(1)D_j(2) \cdots D_j(Q)$ becomes equal to $D_j$, and sends each $D_j(q)$ to $M_q$. Here as an exception $V_j$ discloses $D_j(q)$ only to $M_q$. About $\delta_{j1}, \delta_{j2}, \delta_{j3}$, no one except $V_j$ can know them from $g^{\delta j1}, y_*^{\delta j2}, (gy_*)^{\delta j3}$.
3. $M_1$ calculates pairs $\{g^{sj(1)}, D_j(1)y_*^{sj(1)}\}$ and $\{g^{uj(1)}, r_j(1)y_*^{uj(1)}\}$ and triplet $\{g^{\delta j1 \cdot sj(1)}, y_*^{\delta j2 \cdot sj(1)}, (gy_*)^{\delta j3 \cdot sj(1)}\}$, and sends them to $M_2$.

4. $M_q$ (q > 1) that receives $\{g^{s_{j*}(q-1)}, D_{j*}(q-1)y_*^{s_{j*}(q-1)}\}$, $\{g^{u_{j*}(q-1)}, r_{j*}(q-1)y_*^{u_{j*}(q-1)}\}$ and $\{g^{\delta_{j1} \cdot s_{j*}(q-1)}, y_*^{\delta_{j2} \cdot s_{j*}(q-1)}, (gy_*)^{\delta_{j3} \cdot s_{j*}(q-1)}\}$ from $M_{q-1}$ calculates pairs $\{g^{s_{j*}(q)}, D_{j*}(q)y_*^{s_{j*}(q)}\}$, $\{g^{u_{j*}(q)}, r_{j*}(q)y_*^{u_{j*}(q)}\}$ and triplet $\{g^{\delta_{j1} \cdot s_{j*}(q)}, y_*^{\delta_{j2} \cdot s_{j*}(q)}, (gy_*)^{\delta_{j3} \cdot s_{j*}(q)}\}$, and forwards them to $M_{q+1}$. Where, $s_{j*}(q) = s_j(1)+ \text{---} +s_j(q)$, $u_{j*}(q) = u_j(1)+ \text{---} +u_j(q)$, $D_{j*}(q) = D_j(1)D_j(2) \text{---} D_j(q)$, $r_{j*}(q) = r_j(1)r_j(2) \text{---} r_j(q)$ and $D_{j*}(Q) = D_j$, $r_{j*}(Q) = R_j$.

5. $M_Q$ that calculates pairs $\{g^{s_{j*}(Q)}, D_j y_*^{s_{j*}(Q)}\}$, $\{g^{u_{j*}(Q)}, R_j y_*^{u_{j*}(Q)}\}$ and triplet $\{g^{\delta_{j1} \cdot s_{j*}(Q)} = \mu_{j1}, y_*^{\delta_{j2} \cdot s_{j*}(Q)} = \mu_{j2}, (gy_*)^{\delta_{j3} \cdot s_{j*}(Q)} = \mu_{j3}\}$ in the previous step forwards the pairs to $V_j$ and $M_1$. $M_Q$ sends also $\{\mu_{j1}, \mu_{j2}, \mu_{j3}\}$ to $V_j$.

6. $V_j$ which receives $\{g^{s_{j*}(Q)}, D_j y_*^{s_{j*}(Q)}\}$, $\{g^{u_{j*}(Q)}, R_j y_*^{u_{j*}(Q)}\}$ and $\{\mu_{j1}, \mu_{j2}, \mu_{j3}\}$ confirms that $\{g^{s_{j*}(Q)}, D_j y_*^{s_{j*}(Q)}\}$ is a correct encryption form of $D_j$, i.e. $g^{s_{j*}(Q)}$ and $y_*^{s_{j*}(Q)}$ (= $D_j y_*^{s_{j*}(Q)}/D_j$) in it are calculated as g and $y_*$ to the power of same unknown integer $s_{j*}(Q)$.

7. If encryption form $\{g^{s_{j*}(Q)}, D_j y_*^{s_{j*}(Q)}\}$ is successfully verified, provided that $\Lambda$ is a publicly known integer and $\underline{D}_j(q) = D_j(q)^2$, $\underline{D}_{j*}(q) = (D_j(1)D_j(2) \text{---} D_j(q))^2$ and $\underline{D}_j = D_j^2$, each $M_q$ calculates $\{g^{u_{j*}(Q) \cdot \underline{D}_{j*}(q)}, (R_j y_*^{u_{j*}(Q)})^{\underline{D}_{j*}(q)}\}$ and $\{g^{u_{j*}(Q) \cdot \Lambda \cdot D_{j*}(q)}, (R_j y_*^{u_{j*}(Q)})^{\Lambda \cdot D_{j*}(q)}\}$ from $\{g^{u_{j*}(Q) \cdot \underline{D}_{j*}(q-1)}, (R_j y_*^{u_{j*}(Q)})^{\underline{D}_{j*}(q-1)}\}$ and $\{g^{u_{j*}(Q) \cdot \Lambda \cdot D_{j*}(q-1)}, (R_j y_*^{u_{j*}(Q)})^{\Lambda \cdot D_{j*}(q-1)}\}$ calculated by $M_{q-1}$ to forward them to $M_{q+1}$.

8. $V_j$ calculates $\{g^{u_{j*}(Q) \cdot D_j \cdot (D_j + \Lambda)}, (R_j y_*^{u_{j*}(Q)})^{D_j \cdot (D_j + \Lambda)}\}$ from $\{g^{u_{j*}(Q) \cdot \underline{D}_{j*}(Q)}, (R_j y_*^{u_{j*}(Q)})^{\underline{D}_{j*}(Q)}\}$ and $\{g^{u_{j*}(Q) \cdot \Lambda \cdot D_{j*}(Q)}, (R_j y_*^{u_{j*}(Q)})^{\Lambda \cdot D_{j*}(Q)}\}$ sent by $M_Q$.

9. $V_j$ verifies legitimate calculation of $\{g^{u_{j*}(Q) \cdot D_j \cdot (D_j + \Lambda)}, (R_j y_*^{u_{j*}(Q)})^{D_j \cdot (D_j + \Lambda)}\}$ and constructs initial encryption form $E_0^*(D_j) = <\{g^{s_{j*}(Q)}, D_j y_*^{s_{j*}(Q)}\}, \{g^{u_{j*}(Q)}, R_j y_*^{u_{j*}(Q)}\}, \{g^{u_{j*}(Q) \cdot D_j \cdot (D_j + \Lambda)}, (R_j y_*^{u_{j*}(Q)})^{D_j \cdot (D_j + \Lambda)}\}>$ to put in the encryption stage.

Then, no one can know integer $u_{j*}(Q)$ or $R_j$ unless all mix-servers conspire. Therefore, entities cannot calculate $D_j$ from pair $\{g^{u_{j*}(Q)}, g^{u_{j*}(Q) \cdot D_j \cdot (D_j + \Lambda)}\}$ or $\{R_j y_*^{u_{j*}(Q)}, (R_j y_*^{u_{j*}(Q)})^{D_j \cdot (D_j + \Lambda)}\}$ even if they examine every possible value of $D_j$. In addition, each $D_j(q)$ sent to $M_q$ is a secret of $V_j$ and $M_q$, and as a result, $V_j$ can conceal $D_j$ even from entity *C* that is coercing it if erasable-state voting booths are available. Namely, at a time when *C* asks $V_j$ to disclose $D_j$, $V_j$ can convince *C* that any value *S* is consistent with $E_0^*(D_j)$. Here, an erasable-state voting booth is a one of which memory states are initialized after an entity in it exits. It also disables entities to record the information that they had received and generated in it. This means $V_j$ does not need to reply with the correct value when it is asked about $D_j$ by others.

Nevertheless both $V_j$ and mix-servers can confirm that $E_0^*(D_j)$ finally generated by $V_j$ is legitimate, i.e. $V_j$ verifies them as below and components of $E_0^*(D_j)$ put by $V_j$ were calculated by mix-servers themselves. Although $V_j$ and $M_q$ can construct inconsistent $E_0^*(D_j)$ if they conspire, this dishonesty can be disabled by making $V_j$ anonymous, i.e. among attribute values of other anonymous entities $M_q$ cannot identify $V_j$'s one.

About the verification of $\{g^{s_{j*}(Q)}, D_j y_*^{s_{j*}(Q)}\}$ at step 6, $V_j$ can verify it by confirming relations $g^{s_{j*}(Q) \cdot \delta_{j1}} = \mu_{j1}$, $y_*^{s_{j*}(Q) \cdot \delta_{j2}} = \mu_{j2}$ and $(g^{s_{j*}(Q)} y_*^{s_{j*}(Q)})^{\delta_{j3}} = \mu_{j3}$ through the scheme of Diffie and Hellman. Namely, because discrete logarithm problems are difficult to solve mix-servers that do not know $\delta_{j1}$, $\delta_{j2}$ or $\delta_{j3}$ must calculate $g^{s_{j*}(Q)}$ and $y_*^{s_{j*}(Q)}$ by using same $s_{j*}(Q)$ to satisfy the above relations. Verification of $\{g^{u_{j*}(Q) \cdot D_j \cdot (D_j + \Lambda)}, (R_j y_*^{u_{j*}(Q)})^{D_j \cdot (D_j + \Lambda)}\}$ at step 9 is easy; for $V_j$ that knows $D_j$ and $\Lambda$ it is trivial to confirm that $g^{u_{j*}(Q) \cdot D_j \cdot (D_j + \Lambda)}$ and $(R_j y_*^{u_{j*}(Q)})^{D_j \cdot (D_j + \Lambda)}$ in it are calculated as $g^{u_{j*}(Q)}$ and $R_j y_*^{u_{j*}(Q)}$ to the power of $D_j(D_j + \Lambda)$, i.e. $\{g^{u_{j*}(Q) \cdot D_j \cdot (D_j + \Lambda)}, (R_j y_*^{u_{j*}(Q)})^{D_j \cdot (D_j + \Lambda)}\}$ is a consistent encryption form of $R_j^{D_j \cdot (D_j + \Lambda)}$.

### 3.2. Encryption and Decryption Stages

Mix-servers in the encryption and decryption stages behave in the same way as in Figure 1. Namely, each $M_q$ in the encryption stage encrypts $E_{q-1}^*(D_j) = <\{g^{k_j(q-1)}, D_j y_*^{k_j(q-1)}\}, \{g^{v_j(q-1)}, R_j y_*^{v_j(q-1)}\}, \{g^{w_j(q-1)}, R_j^{D_j \cdot (D_j + \Lambda)} y_*^{w_j(q-1)}\}>$ received from $M_{q-1}$ to $E_q^*(D_j) = <\{g^{k_j(q)}, D_j y_*^{k_j(q)}\}, \{g^{v_j(q)}, R_j y_*^{v_j(q)}\}, \{g^{w_j(q)}, R_j^{D_j \cdot (D_j + \Lambda)} y_*^{w_j(q)}\}>$. And $M_q$ in the decryption stage receives $F_q^*(D_j) = <\{g^{k_j(q)}, D_j g^{k_j(Q) \cdot x^*(q)}\}, \{g^{v_j(q)}, R_j g^{v_j(Q) \cdot x^*(q)}\}, \{g^{w_j(q)}, R_j^{D_j \cdot (D_j + \Lambda)} g^{w_j(q) \cdot x^*(q)}\}>$ from $M_{q+1}$, and while using decryption key x(q) decrypts it to $F_{q-1}^*(D_j) = <\{g^{k_j(Q)}, D_j g^{k_j(Q) \cdot x^*(q-1)}\}, \{g^{v_j(Q)}, R_j g^{v_j(Q) \cdot x^*(q-1)}\}, \{g^{w_j(Q)}, R_j^{D_j \cdot (D_j + \Lambda)} g^{w_j(q) \cdot x^*(q-1)}\}>$. Then, $M_1$ finally decrypts $F_1^*(D_j)$ to $F_0^*(D_j) = <\{g^{k_j(Q)}, D_j\}, \{g^{v_j(Q)}, R_j\}, \{g^{w_j(Q)}, R_j^{D_j \cdot (D_j + \Lambda)}\}>$, and convinces others that $F_0^*(D_j)$ is legitimate by pair $\{R_j, R_j^{D_j \cdot (D_j + \Lambda)}\}$. Here, $k_j(q)$, $v_j(q)$ and $w_j(q)$ are secret integers of $M_q$, $k_{j*}(q) = s_{j*}(Q)+k_j(1)+k_j(2)+ \text{---} +k_j(q)$, $v_{j*}(q) = u_{j*}(Q)+v_j(1)+v_j(2)+ \text{---} +v_j(q)$, $w_{j*}(q) = u_{j*}(Q)D_j(D_j+\Lambda)+w_j(1)+w_j(2)+ \text{---} +w_j(q)$ and $x_*(q) = x(1)+x(2)+ \text{---} +x(q)$.

### 3.3. Verifying Behaviors of Revised-SVRM

About illegitimate behaviors in the revised-SVRM, $R_j^{D_j \cdot (D_j + \Lambda)}$ in the 3rd term in $E_q^*(D_j)$ means that final decryption result $F_0^*(D_j) = <\{g^{k^*(Q)}, \alpha\}, \{g^{v^*(Q)}, \beta\}, \{g^{w^*(Q)}, \gamma\}>$ must satisfy relation $\gamma = \beta^{\alpha \cdot (\alpha + \Lambda)}$. By using this relation, although each encryption form $E_q^*(D_j)$ differs from $E_q(D_j)$, illegitimate behaviors of revised-SVRM can be detected and liable entities can be identified as same as in the original SVRM. But to verify behaviors in the decryption stage, $M_q$ must disclose also $\sigma_q = s_1(q)+s_2(q)+ \text{---} +s_N(q)$ in addition to $\kappa_q = k_1(q)+k_2(q)+ \text{---} +k_N(q)$ because $G_d$ in Sec. 3 is calculated as $y_*^{k1*(Q)+ \text{---} +KN*(Q)}$.

When compared with the original SVRM, identification of dishonest entities becomes simpler. Namely, because $V_j$ and mix-servers in the unknown number generation stage had mutually confirmed their legitimate behaviors already, examination of behaviors in the unknown number generation stage is not necessary.

## 4. Revised-SVRM Based Voting Scheme

This section develops a voting scheme while exploiting revised-SVRM, anonymous tag based credentials and erasable-state voting booths. The scheme consists of voters $V_1, V_2, \text{---}, V_N$, election authority *A* and mix-servers $M_1, M_2, \text{---}, M_Q$ in the encryption, decryption and unknown number generation stages. Elections proceed through the voter registration, voting, pre-tallying and tallying phases as below.

### 4.1 Voter Registration

Firstly, each voter $V_j$ shows its identity to election authority *A* at an entrance of an election site. After that, *A* gives credential $T(A, V_j, Z_j)$ to $V_j$ if it is eligible, and in exchange for the credential $V_j$ issues a receipt to *A*.

Then, because $A$ knows who is $V_j$, $V_j$ cannot obtain multiple credentials. On the other hand, the receipt ensures that $V_j$ certainly can obtain its credential, i.e. $A$ must show the receipt issued by $V_j$ to reject $V_j$'s request.

## 4.2. Voting

### 4.2.1. Entering a Voting Booth

Each voter $V_j$ shows its credential $T(A, V_j, Z_j)$ to authority $A$ and calculates used seal $U_0^{Zj}$ of the credential from publicly known integer $U_0$ defined by $A$, and if $T(A, V_j, Z_j)$ is legitimate and $U_0^{Zj}$ was not calculated before, $V_j$ is allowed to enter its choosing voting booth. Then features of credentials and used seals allow only eligible voters to enter voting booths only once.

### 4.2.2. Vote Construction

In the voting booth, $V_j$ constructs pair $E_0^*(D_j) = <\{g^{s_{j*}(Q)}, D_j y_*^{s_{j*}(Q)}\}, \{g^{u_{j*}(Q)}, R_j y_*^{u_{j*}(Q)}\}, \{g^{u_{j*}(Q) \cdot D_j(D_j+\Lambda)}, (R_j y_*^{u_{j*}(Q)})^{D_j(D_j+\Lambda)}\}>$ and $E_0^*(D_j, \Omega) = <\{g^{s_{j*}(Q)}, \Gamma_j y_*^{s_{j*}(Q)}\}, \{g^{u_{j*}(Q)}, R_j y_*^{u_{j*}(Q)}\}, \{g^{u_{j*}(Q) \cdot \Gamma_j(\Gamma_j+\Lambda)}, (R_j y_*^{u_{j*}(Q)})^{\Gamma_j(\Gamma_j+\Lambda)}\}>$ as an initial encryption form of its vote, and forwards it to mix-server $M_1$ in the encryption stage. Where, $\Lambda$ is a publicly known constant integer, $R_j$ is an integer no one knows and $D_j$ represents a candidate $V_j$ chooses. Also, provided that each $s_j(q)$, $u_j(q)$, $\underline{s}_j(q)$ and $\Omega(q)$ are $M_q$'s secret integers, $s_{j*}(Q) = s_j(1)+ --- +s_j(Q)$, $u_{j*}(Q) = u_j(1)+ --- +u_j(Q)$, $\underline{s}_{j*}(Q) = \underline{s}_j(1)+ --- +\underline{s}_j(Q)$, $\Omega = \Omega(1)\Omega(2) --- \Omega(Q)$ and $\Gamma_j = D_j^{\Omega}$.

In detail, $V_j$ decomposes $D_j$ into $\{D_j(1), D_j(2), ---, D_j(Q)\}$ so that $D_j = D_j(1)D_j(2) --- D_j(Q)$ holds, and informs each mix-server $M_q$ in the unknown number generation stage of $D_j(q)$. After that, jointly with the mix-servers $V_j$ calculates $E_0^*(D_j)$ as in Sec. 3. About $E_0^*(D_j, \Omega)$, firstly each $M_q$ generates its secret integer $\Omega(q)$ and calculates $D_j(q)^{\Omega(q)}$ to forward it to $M_{q+1}$. Then, $M_{q*}$ that receives $D_j(q)^{\Omega(q) \cdot \Omega(q+1) --- \Omega(q*-1)}$ from $M_{q*-1}$ calculates $D_j(q)^{\Omega(q) \cdot \Omega(q+1) --- \Omega(q*-1) \cdot \Omega(q*)}$, and sends it to $M_{q*+1}$ ($M_1$ is regarded as $M_{Q+1}$). As a result, $M_q$ receives $D_j(q)^{\Omega} = \Gamma_j(q)$ from $M_{q-1}$, and $V_j$ and mix-servers can calculate $E_0^*(D_j, \Omega)$ from values $\Gamma_j(1), \Gamma_j(2), ---, \Gamma_j(Q)$ as they calculated $E_0^*(D_j)$.

In the above, it must be noted that no one knows the value of $\Omega$ because only $M_q$ knows $\Omega(q)$. Also, $V_j$ can verify legitimate calculation of each $\Gamma_j(q)$ through the scheme of Diffie and Hellman by asking mix-servers to calculate $D_j(q)^{\Phi \cdot \Omega}$ from $D_j(q)^{\Phi}$ ($\Phi$ is $V_j$'s secret integer).

### 4.2.3. Vote Approval

Before exiting the voting booth, $V_j$ approves that pair $\{E_0^*(D_j), E_0^*(D_j, \Omega)\}$ put in the encryption stage is legitimate. Namely, after confirming that the pair disclosed by $M_1$ is correct, $V_j$ calculates another used seal $U_1^{Zj}$ of $T(A, V_j, Z_j)$ from publicly known integer $U_1$ defined by $A$, and $A$ discloses it publicly with $U_0^{Zj}$. Here, used seal $U_1^{Zj}$ is $V_j$'s approval of pair $\{E_0^*(D_j), E_0^*(D_j, \Omega)\}$, i.e. $V_j$ can replace it with new ones until it discloses $U_1^{Zj}$. On the other hand, because only $V_j$ can calculate $U_1^{Zj}$, $A$ can reject $V_j$'s requests about replacements of the pair after $U_1^{Zj}$ is disclosed.

### 4.2.4. Vote Encryption

Once, $\{E_0^*(D_j), E_0^*(D_j, \Omega)\}$ is approved, mix-servers in the encryption stage encrypt it to $E_Q^*(D_j) = <\{g^{k_{j*}(Q)}, D_j y_*^{k_{j*}(Q)}\}, \{g^{v_{j*}(Q)}, R_j y_*^{v_{j*}(Q)}\}, \{g^{w_{j*}(Q)}, R_j^{D_j(D_j+\Lambda)} y_*^{w_{j*}(Q)}\}>$ and $E_Q^*(D_j, \Omega) = <\{g^{k_{j*}(Q)}, \Gamma_j y_*^{k_{j*}(Q)}\}, \{g^{v_{j*}(Q)}, R_j y_*^{v_{j*}(Q)}\}, \{g^{w_{j*}(Q)}, R_j^{\Gamma_j(\Gamma_j+\Lambda)} y_*^{w_{j*}(Q)}\}>$ while shuffling their encryption results. Where, provided that $k_j(q)$, $v_j(q)$, $w_j(q)$, $\underline{k}_j(q)$, $\underline{v}_j(q)$ and $\underline{w}_j(q)$ are $M_q$'s secret integers, $k_{j*}(q) = s_j(Q)+k_j(1)+ --- +k_j(q)$, $v_{j*}(q) = u_{j*}(Q)+v_j(1)+ --- +v_j(q)$, $w_{j*}(q) = u_{j*}(Q)D_j(D_j+\Lambda)+w_j(1)+ --- +w_j(q)$, $\underline{k}_{j*}(q) = \underline{s}_{j*}(Q)+\underline{k}_j(1)+ --- +\underline{k}_j(q)$, $\underline{v}_{j*}(q) = u_{j*}(Q)+\underline{v}_j(1)+ --- +\underline{v}_j(q)$ and $\underline{w}_{j*}(q) = u_{j*}(Q)\Gamma_j(\Gamma_j+\Lambda)+\underline{w}_j(1)+ --- +\underline{w}_j(q)$.

## 4.3. Pre-tallying

Votes encrypted in the voting phase are decrypted by mix-servers $M_Q$, $M_{Q-1}$, ---, $M_1$ in the decryption stage, but they decrypt only selected ones. The pre-tallying phase determines these votes. Namely, mix-servers in this phase decrypt only $E_Q^*(D_j, \Omega)$ in each pair $\{E_Q^*(D_j), E_Q^*(D_j, \Omega)\}$, and as a result only decryption form $F_0^*(D_j, \Omega) = <\{g^{k_{j*}(Q)}, \Gamma_j = D_j^{\Omega}\}, \{g^{v_{j*}(Q)}, R_j\}, \{g^{w_{j*}(Q)}, R_j^{\Gamma_j(\Gamma_j+\Lambda)}\}>$ is disclosed.

Then, election authority $A$ compares disclosed $D_1^{\Omega}$, $D_2^{\Omega}$, ---, $D_N^{\Omega}$, and determines $E_Q^*(D_h)$ that corresponds to $E_Q^*(D_h, \Omega_*(Q))$ is an inferior vote that will not be decrypted when value $D_h^{\Omega}$ appears less than the predefined number of times in set $\{D_1^{\Omega}, D_2^{\Omega}, ---, D_N^{\Omega}\}$. Here, because no one knows integer $\Omega$ anyone cannot know $D_j$ from $D_j^{\Omega}$. Decryption of $E_Q^*(D_j, \Omega)$ itself is carried out totally in the same way as in Sec. 3.

## 4.4. Tallying

Because authority $A$ can determine election winners without decrypting inferior votes, mix-servers in the tallying phase decrypt encryption form $E_Q^*(D_j)$ only when it corresponds to a non-inferior vote. As a result, voters can conceal correspondences between them and their votes from others even when they are forced to choose candidates unique to them.

As same as $E_Q^*(D_j, \Omega)$, decryption of $E_Q^*(D_j)$ is carried out as in Sec. 3. But mix-servers do not need to decrypt all non-inferior votes because $E_Q^*(D_j)$ and $E_Q^*(D_h)$ are decrypted to same value $D_j$ if $E_Q^*(D_j, \Omega)$ and $E_Q^*(D_h, \Omega)$ were decrypted to $D_j^{\Omega}$. Therefore, mix-servers decrypt only 1 encryption form $E_Q^*(D_j)$ from a set of encryption forms that are accompanied by same value $D_j^{\Omega}$.

## 4.5. Detecting Dishonesties and Identifying Liable Entities

Revised-SVRM used in the developed e-voting scheme enables any entity to detect illegitimately handled votes efficiently as discussed in Sec. 3. It also enables election authority $A$ to identify entities liable for dishonesties without revealing votes of honest voters.

In addition, in cases when mix-servers are determined dishonest, $A$ can force them to correctly reprocess illegitimately handled votes. Namely, because voters and mix-servers in the unknown number generation stage had verified their behaviors mutually, initial encryption forms put in the encryption stage are ensured to be legitimate. Then, once all voters had approved initial encryption forms of their votes, $A$ and mix-servers can reprocess illegitimately handled votes until their decryption results become consistent without reelections.

In the above, even if voter $V_j$ and mix-server $M_q$ in the unknown number generation stage conspire, they cannot put an inconsistent initial encryption form. The reason is that $V_j$ is anonymous and $M_q$ cannot identify $V_j$'s vote. To handle $V_j$'s vote illegitimately $M_q$ must take a risk that its dishonesty is revealed, i.e. $V_h$ claims $M_q$ is dishonest if $M_q$ generates an initial encryption form of $V_h$'s vote inconsistently instead of $V_j$'s one. In the same way, $V_j$ can protect itself from threats where conspiring mix-server $M_1$ and entity $C$ that coerces $V_j$ know $V_j$'s vote. In detail, when $M_1$ in the encryption stage encrypts initial encryption form $\{E_0^*(D_j), E_0^*(D_j, \Omega)\}$ of $V_j$ inconsistently, the dishonest entity identification procedure reveals the correspondence between final decryption form $\{F_0^*(D_j), F_0^*(D_j, \Omega)\}$ and $\{E_0^*(D_j), E_0^*(D_j, \Omega)\}$. But $M_1$ cannot identify $\{E_0^*(D_j), E_0^*(D_j, \Omega)\}$ because $V_j$ is anonymous.

## 5. Features of the Developed Scheme

The e-voting scheme developed based on revised-SVRM satisfies all essential requirements of elections as follow.

*Privacy*  As discussed in Sec. 3 and 5, no one except voter $V_j$ itself can know candidate $D_j$ that $V_j$ had chosen. But $V_j$ that did not register itself cannot conceal its abstention because voters register themselves by showing their identities. To conceal its abstention from others $V_j$ must register itself and put an invalid vote or leave the election site without entering a voting booth.

*Verifiability*  Anonymous credential ensures that only eligible entities can put votes, and used seals of credentials disable voters to put votes multiple times. About tallying, all votes put by voters and vote forms handled by mix-servers are publicly disclosed and revised-SVRM is verifiable. Then, anyone including third parties can verify the accuracy of elections.

*Fairness*  No one can know the interim election results because the scheme does not disclose votes in their plain forms until the end of the pre-tallying phase.

*Incoercibility*  Voter $V_j$ can conceal candidate $D_j$ in $\{E_0^*(D_j), E_0^*(D_j, \Omega)\}$ from $C$ that is coercing it, i.e. because $D_j$ is encrypted by using unknown integers, $V_j$ can declare that $\{E_0^*(D_j), E_0^*(D_j, \Omega)\}$ is an encryption form of any candidate $S$. Also, erasable-state voting booths disable $C$ to obtain enough information from $V_j$ to reconstruct $D_j$ even if $C$ is conspiring with several mix-servers. Because inferior votes are not decrypted, $C$ cannot confirm whether $V_j$ had chosen $C$'s designating candidate $S$ or not even when $S$ is unique to $V_j$.

Here because $V_j$ is anonymous, as discussed at the end of Sec. 4.5, $C$ cannot know the correspondence between initial encryption form $\{E_0^*(D_j), E_0^*(D_j, \Omega)\}$ and final decryption result $\{F_0^*(D_j), F_0^*(D_j, \Omega)\}$ even if it conspires with 1st mix-server $M_1$ in the encryption stage or mix-servers in the unknown number generation stage. In detail, $A$ in the voter registration phase gives credential $T(A, V_j, Z_j)$ to $V_j$ just before $V_j$ enters a voting booth, therefore $V_j$ cannot inform $C$ or mix-servers of integer $Z_j$ in $T(A, V_j, Z_j)$ so that they can identify $V_j$'s vote.

But it must be noted that $C$ which is forcing $V_j$ to abstain from the election can confirm whether $V_j$ actually had abstained or not by asking $V_j$ to recalculate the used seal $V_j$ had calculated in the voter registration phase. This threat exists also in usual paper based elections, and currently an only way to remove this threat is to introduce regulations that force all voters to visit election sites regardless that they choose valid candidates or not.

Even if election authority $A$ gives 2 anonymous credentials $T_\alpha$ and $T_\beta$ to $V_j$, $C$ can know whether $V_j$ actually had abstain or not. Namely, although $V_j$ can visit an election site without revealing its identity by showing $T_\beta$ (where $V_j$ obtains $T_\beta$ by showing $T_\alpha$ that it had obtained in advance while showing its identity), $C$ can know $V_j$ even from $T_\beta$ if it asks $V_j$ to disclose secrets in $T_\beta$.

*Robustness*  Because initial encryption form $\{E_0^*(D_j), E_0^*(D_j, \Omega)\}$ of a vote put by voter $V_j$ is verified by mix-servers and $V_j$ itself, $V_j$ cannot claim that mix-servers had constructed it illegitimately. Therefore, once encrypted votes are successfully disclosed, revised-SVRM enables reprocessing of votes until final decryption forms are disclosed correctly without reelections.

## 6. Conclusion

Based on revised-SVRM an e-voting scheme that satisfies all essential requirements of elections was developed, i.e. it satisfies requirements about privacy, verifiability, fairness, incoercibility and robustness. But the scheme assumes state-erasable voting booths. Therefore as one of future works, efficient schemes for implementing state-erasable voting booths must be developed.

## References

[1] Diffie and M. E. Hellman, "New directions in cryptography," *IEEE Trans. On Information Theory*, IT-22(6), 644-654, 1976.

[2] D. Boneh and P. Golle, "Almost entirely correct mixing with applications to voting," *ACM Conference on Computer and Communication Security*, 68-77, 2002.

[3] P. Golle, S. Zhong, D. Boneh, M. Jakobsson and A. Juels, "Optimistic mixing for exit-polls," *Asiacrypt 2002*, 451-465, 2002.

[4] M. Jakobson, A. Juels and R. Rivest, "Making mix nets robust for electronic voting by randomized partial checking," *USENIX Security '02*, 339-353, 2002.

[5] L. Nguen, R. Dafavi-Naini and K. Kurosawa, "Verifiable shuffles: A formal model and a Paillier-based efficient construction with provable security," *PKC 2004*, 61-75, 2004.

[6] B. Lee, C. Boyd, E. Dawson, K. Kim, J. Yang and S. Yoo, "Providing receipt-freeness in mixnet-based voting protocols," *Proceedings of the ICISC '03*, 261-74, 2003.

[7] J. Furukawa, "Efficient, Verifiable shuffle decryption and its requirement of unlinkability," *PKC 2004*, 319-332, 2004.

[8] K. Sampigethaya and R. Poovendran, "A framework and taxonomy for comparison of electronic voting schemes," *Elsevier Computers and Security*, 25, 137-153, 2006.

[9] S. Weber, "A coercion-resistant cryptographic voting protocol - evaluation and prototype implementation," *Diploma thesis, Darmstadt University of Technology*; 2006

[10] P. Y. A. Ryan, D. Bismark, J. Heather, S.Schneider and Z. Xia, "A voter verifiable voting system," *IEEE Trans. On Information Forensics and Security*, 4(4), 662-673, 2009.

[11] K. A. Md Rokibul, S. Tamura, S. Taniguchi and T. Yanase, "An anonymous voting scheme based on confirmation numbers," *IEEJ Trans. EIS*. 130(11), 2065-2073, 2010.

[12] C. C. Lee, T. Y. Chen, S. C. Lin and M. S. Hwang, "A new proxy electronic voting scheme based on proxy signatures," *Lecture Notes in Electrical Engineering*, 164, Part 1 3-12, 2012.

[13] S. Tamura, "Anonymous security systems and applications: requirements and solutions," *Information Science Reference*, 2012.

[14] S. Tamura and S. Taniguchi, "Simplified verifiable re-encryption mix-nets," *Information Security and Computer Fraud*, 1(1), 1-7, 2013.

[15] S. Tamura and S. Taniguchi, "Enhancement of anonymous tag based credentials," *Information Security and Computer Fraud*, 2(1), 10-20, 2014.

[16] S. Tamura, "Elements of schemes for preserving privacies in e-society systems," *Lambert Academic Publishing*, 2015.